\documentclass[twocolumn,aps,superscriptaddress,showpacs]{revtex4}

\usepackage{amsmath}
\usepackage{amssymb}
\usepackage{amsfonts}
\usepackage{amsthm}
\usepackage{mathrsfs}

\usepackage{color}

\def\A{\mathcal{A}}
\def\B{\mathcal{B}}
\def\C{\mathcal{C}}
\def\F{\mathcal{F}}
\def\K{\mathcal{K}}
\def\Epsilon{\mathrm{E}}
\def\Kappa{\mathrm{K}}

\def\PBG#1#2{\{#1, #2\}^G}
\def\PBGamma#1#2{\{#1, #2\}^\Gamma}
\def\PBK#1#2{\{#1, #2\}^K}
\def\PBA#1#2{\{#1, #2\}^A}
\def\PBKappa#1#2{\{#1, #2\}^{\Kappa}}
\def\PBAlpha#1#2{\{#1, #2\}^\Alpha}
\def\d{\mathrm{d}}
\def\fd#1#2{\frac{\delta#1}{\delta#2}} % functional derivative
\def\H{\mathcal{H}}

\def\Alpha{\mathrm{A}}

\begin{document}

\title{Unambiguous spin-gauge formulation of canonical general relativity with conformorphism invariance}

\author{Charles H.-T. Wang}
%\email{c.wang@abdn.ac.uk}
\affiliation{School of Engineering and Physical Sciences,
University of Aberdeen, King's College, Aberdeen AB24 3UE, Scotland}
\affiliation{Space Science and Technology Department,
Rutherford Appleton Laboratory, Didcot, Oxon OX11 0QX, England}

%\date{\today}
%\date{10 July 2005}

%
%{test text}
%

\begin{abstract}
We present a parameter-free gauge formulation of general
relativity in terms of a new set of real spin connection
variables. The theory is constructed by extending the phase space
of the recently formulated conformal geometrodynamics for
canonical gravity to accommodate a spin gauge description. This
leads to a further enlarged set of first class gravitational
constraints consisting of a reduced Hamiltonian constraint and the
canonical generators for spin gauge and conformorphism
transformations. Owing to the incorporated conformal symmetry,
the new theory is shown to be free from an ambiguity of the
Barbero-Immirzi type.
\end{abstract}


\maketitle

%\textbf{Introduction.}

The success of the gauge field theory has long inspired efforts to
reformulate Einstein's general relativity (GR) into a canonically
quantizable form resembling that of Yang-Mills.
A major milestone
in this regard is Ashtekar's introduction of a set of ``new''
variables for GR \cite{Ashtekar1986, Ashtekar1987}. In essence,
they consist of a spin connection on the spatial hypersurface
with the densitized triad as conjugate momentum. The extrinsic
curvature enters into the spin connection as a torsion term, so
that the spin connection is independent of its momentum. This is
the case since the densitized triad uniquely specifies the
intrinsic geometry which in turn determines the
torsion-free, i.e. Levi-Civita (LC), spin connection.

Following Sen's pioneering work \cite{Sen}, the coefficient of
this torsion term is chosen to be an imaginary number
corresponding to the use of SL$(2,\mathbb{C})$ as the gauge group.
This choice has the advantage of simplifying the gravitational
constraints to a polynomial form that permits a close analogy with
the Hamiltonian of the Yang-Mills theory, and has been
considered as a crucial rationale behind Ashtekar's original
complex approach to the gauge formulation of GR.
This approach has grown into the active research area of ``loop quantum
gravity'' using non-perturbative quantization techniques, as
envisaged by Gambini, Jacobson, Rovelli, Smolin \textit{et al}
\cite{JacobsonSmolin1988, GambiniPullin1996, Rovelli1998, Thiemann2001, AshtekarLewandowski2004}.

However, the utility of such an approach has so far been limited
mainly  to solving the momentum constraints. Problems such as the
physical inner product and the meaning of time \cite{Isham1993}
are left unaddressed. Furthermore, the imposition of certain
reality conditions for Ashtekar's complex variables to yield real
observables has proven to be problematic. This has prompted
Barbero to consider an alternative set of spin gauge variables
\cite{Barbero1995a} based on the real spin gauge group SO$(3)$. A
spin connection is also employed, with an arbitrary real positive
parameter entering as the coefficient of the torsion term. Immirzi
noted that the appearance of this free (Barbero-Immirzi) parameter
in the resulting Hamiltonian constraint makes the quantized theory
ambiguous \cite{Immirzi1997}. Although ongoing efforts are being
made to fix this  ambiguity by tuning the Barbero-Immirzi
parameter to match certain quantum black-hole entropy predictions
with the classical results using the Hawking-Bekenstein formula
\cite{Ambiguity}, such a state of the affair seems unsatisfactory
from a theoretical viewpoint.

Following a previous paper \cite{Wang2005b}, we show in this work
how a real and parameter-free gauge formulation of GR can arise
from an extended phase space of canonical gravity that
incorporates conformal as well as spin symmetry. In what
follows, the canonical analysis is rooted in the
Arnowitt-Deser-Misner (ADM) formulation with metric signature
$(-,+,+,+)$ and a compact spatial sector. Units of
$16 \pi G = c = \hbar = 1$ are adopted.
%The convention of spin indices follows mainly that of
%\cite{Thiemann2001, AshtekarLewandowski2004}.

\textbf{Conformal geometrodynamics.}
The standard geometrodynamics is based on the ADM
variables of GR consisting of the spatial metric
$g_{a b}$ with conjugate momentum
\begin{equation}\label{pK^ab}
p^{ab}
=
\mu (K^{ab} - g^{ab} K)
\end{equation}
in terms of
the (volume) scale factor
$\mu := \sqrt{\det g_{ab}}$
and
the extrinsic curvature tensor is
$K_{a b} = (2N)^{-1}[\dot{g}_{a b} - 2 \nabla_{(a} N_{b)}]$
where the overdot denotes a time-derivative, $N$ is
the lapse function  and $N^a$
the shift vector. The trace $K = g^{a b} K_{a b}$ is the mean
(extrinsic) curvature. We denote the LC connection
of $g_{a b}$ by $\nabla$ and will also use a subscript ``$;$''
for the associated covariant differentiation.
The ADM momentum and Hamiltonian constraints are given
respectively by
\begin{align}
\H_a &= -2{p}^b{}_{a;b}
\\
\H_\perp &= G_{a b c d}\, p^{a b} p^{c d} -  \mu R
\end{align}
where
$G_{a b c d} := (2\mu)^{-1} (g_{a c}g_{b d} + g_{a d}g_{b c} -  g_{a b}g_{c d})$
and
$R := R^{a b}{}_{a b}$ is the Ricci scalar curvature of $g_{ab}$ in terms
of the Riemann curvature tensor $R^a{}_{b c d}$ satisfying
$[\nabla_c, \nabla_d] V^a = R^a{}_{b c d} V^b$ for any vector $V^a$.
We shall analyze a sequence of canonical transformations
of the gravitational variables. To keep track of different
variables being used at each stage, we shall attribute them to
a capital letter. We start by referring to $(g_{a b}, p^{a b})$ as the ``$G$-variables'',
w.r.t. which the Poisson bracket (PB) is denoted by
$\PBG{\centerdot\,}{\centerdot}$.

In \cite{Wang2005b},
the problem of time and true dynamical degrees of freedom of GR has been
reexamined in the canonical framework
by extending the ADM phase space to that
of York's mean curvature time $\tau := (4/3) K$ with $\mu$ as momentum and conformal
metric $\gamma_{a b}$ with momentum $\pi^{ab}$.
Based on York's decomposition of tensors \cite{York},
a canonical transformation has been found
to relate the $G$-variables to
the ``$\Gamma$-variables'' $(\gamma_{a b}, \pi^{a b}; \tau, \mu)$
via
\begin{align}\label{ggamma}
{g}_{a b}
&=
\phi^4 \gamma_{a b},
\;\;
%\label{pexpr}
p^{a b}
=
\phi^{-4}\pi^{a b}
-
\frac12  \, \phi^2 \bar{\mu}\, \gamma^{ab}\tau .
\end{align}
Here $\bar{\mu} := \sqrt{\det \gamma_{ab}}$
is the
conformal scale factor and
$\phi := ({\mu}/{\bar{\mu}})^{1/6}$
is the conformal factor. The transformation
using \eqref{ggamma} is canonical
since the original canonical PB relations for the $G$-variables
are strongly preserved by
the
PBs w.r.t. the $\Gamma$-variables denoted by
$\PBGamma{\centerdot\,}{\centerdot}$. Using an arbitrary
function $f$, the
local rescaling
$\gamma_{a b}\rightarrow f^4 \gamma_{a b}, \pi^{a b}\rightarrow f^{-4} \pi^{a b}$,
while holding $\tau$ and $\mu$ unchanged,
leave ${g}_{a b}, p^{a b}$ invariant. This redundancy of the $\Gamma$-variables
is offset by introducing the ``conformal constraint'':
\begin{equation}\label{conformconst}
\C^\Gamma
:=
{\gamma}_{ab}\pi^{ab}
\end{equation}
which generates local rescaling
transformations through its PBs
with $\gamma_{a b}, \pi^{a b}, \tau$ and $\mu$ \cite{Wang2005b}.
In terms of $\C^\Gamma$, we have
\begin{align}\label{diffconst}
\H_a
&=
\C_a^\Gamma
+4(\ln\phi)_{,a}\, \C^\Gamma
\\
\label{C^Gamma_a}
\C_a^\Gamma
&:=
\tau_{,a} {\mu} -2 \pi^b{}_{a;b}
\\
\H_\perp
&=
\C_\perp^\Gamma
+
\frac\tau2\,\C^\Gamma
-
\frac1{2\mu}\,(\C^\Gamma)^2
\label{HR}
\\
\C_\perp^\Gamma
&:=
-\frac{3}{8}\, \tau^2 \mu
+
\frac1{\mu}\,\pi{}_{ab}\pi^{ab}
-\mu R\label{H^Gamma} .
\end{align}
In the $\Gamma$-variables, $\C^\Gamma_a$
play the role of the diffeomorphism
constraints. The PBs amongst $\C^\Gamma_a$
and $\C^\Gamma$ satisfy the Lie algebra for
conformorphisms on the spatial hypersurface \cite{Wang2005b, York, FischerMarsden1977}.
It follows that
$\{\C^\Gamma, \C_a^\Gamma, \C_\perp^\Gamma\}$ form a set of independent
first class constraints, by noting
the preservation
$\PBG{\H_\perp(x)}{\H_\perp(x')}=\PBGamma{\H_\perp(x)}{\H_\perp(x')}$
and that all summands in \eqref{diffconst} and \eqref{HR}
are function-proportional to these constraints.

\textbf{Triad formalism.} Let us take one step
of the canonical analysis
back to the $G$-variables.
Introduce the triad $e^i_a$ with inverse $e_i^a$
and densitized triad $E_i^a = \mu\, e_i^a$ with inverse
$E_i^a$,
so that% the metric has the form:
\begin{equation}\label{gabE}
g_{a b} = \mu^2  E^i_a  E^i_b
\end{equation}
with its inverse given by $g^{a b} = \mu^{-2} E^a_i E^b_i$,
using the ``spin indices'' $i, j, \cdots = 1, 2, 3$. We
choose the orientation so that $\det e^i_a = \mu > 0$.
Introduce the spin-valued extrinsic curvature $K^i_{a}$ so that
the extrinsic curvature tensor is
\begin{equation}\label{Kab1}
{K}_{ab}
=
\frac\mu2 \, {K}^i_{(a} E^i_{b)} .
\end{equation}
The ``$K$-variables'' $(K^i_{a}, E_i^a)$ are coordinates of an
extended phase space of GR. The redundancy is due to the spin
transformation and can be eliminated by means of the constraint:
\begin{equation}\label{gauss0}
\K_{a b}
:=
\mu^2 {K}^i_{[a} E^i_{b]} .
\end{equation}
The $K$-variables are
canonical since by regarding the $G$-variables
as functions of them via
\eqref{pK^ab} and \eqref{gabE} we have
\begin{align}\label{PBgpKE}
% QG050101a
\PBK { g_{a b}(x)}{ p^{c d}(x') }
&=
\delta^{c d}_{a b}\, \delta(x, x')
\\
\label{PBggKE}
%easy
\PBK  {g_{a b}(x)}  {g_{c d}(x')}
&=
0
\\
\label{PBppKE}
% QG050101
\PBK { p^{a b}(x)} {p^{c d}(x')}
&=
\K^{a b c d}(x)\,\delta(x, x')
\end{align}
in terms of
$\K^{a b c d}:=1/8\,(g^{ac} \K^{b d}+g^{ad} \K^{b c}+g^{bc} \K^{a d}+g^{bd} \K^{a c})$ \cite{Ashtekar1987}.
From \eqref{PBgpKE}--\eqref{PBppKE}, we have in general
\begin{equation*}
\PBK \A \B
=
\PBG \A \B
+
\int
\fd{\A}{p^{a b}(x)}\fd{\B}{p^{c d}(x)}
\K^{a b c d}(x)
\d^3 x .
%\label{ABKE}
\end{equation*}
However, by exploiting the property $\K^{a b c d} = \K^{(a b) (c d)}=-\K^{(c d) (a b)}$
and algebraic dependence of $\H$ on $p^{a b}$ we get
\begin{equation}\label{PBKGHH}
\PBK{\H(x)}{\H(x')}=\PBG{\H(x)}{\H(x')}
\end{equation}
which will prove useful in simplifying our canonical analysis tasks.
Instead of $\K_{a b}$, it is advantageous to adopt
\begin{equation}\label{gauss2}
\C_i^K
:=
\epsilon^{}_{i j k} K^{}_{a j} E^a_{k}
=
-\frac{1}{\mu^2} \epsilon^{}_{i j k} \K_{a b} E_j^a E_k^b
\end{equation}
serving as the canonical generator for rotation (spin).
%via
%\begin{align}\label{PBGiKjc}
%\PBK{\C_{i}^K(x)}{K^j_c(x')} &= \epsilon^{}_{ijk} K^{k}_{c}(x)\, \delta(x, x')
%\\
%\label{PBGiEjc}
%\PBK{\C_{i}^K(x)}{E_j^c(x')} &= \epsilon^{}_{ijk} E^c_{k}(x)\, \delta(x, x')
%\end{align}
A natural connection associated with the spin indices is the LC
spin connection $\Gamma^i_a$, so that the associated spin
covariant derivative of any spin-valued scalar $S^i$ is given by
\begin{equation}\label{spindel}
\nabla^{}_a S^i = \partial^{}_a S^i + \epsilon^{}_{i j k} \Gamma^j_a S^k .
\end{equation}
The spin connection itself is uniquely determined
in terms of $E_i^a$
by the torsion-free
condition
\begin{equation}\label{torfree}
\nabla^{}_c E_i^a
=
0 .
\end{equation}
% For the sake of notational economy,
% symbol overloading is used heavily -- the meaning of an indexed same depends on its indexing structure.
By varying \eqref{torfree} we see a relation of the form
\begin{equation}\label{var_LC_spin_connex}
\delta \Gamma_i^a
=
E^{a c}_{b i j}
\delta E^b_{j;c}
\end{equation}
where $E^{a c}_{b i j}$ is algebraic in $E_i^a$ and
so $\nabla^{}_d E^{b c}_{a i j} = 0$.
This form has two remarkable consequences:
First, for any scalar $\varphi$ satisfying $\d\varphi=0$,
if $\delta E_i^a = E_i^a \delta\varphi$  then
$\delta\Gamma_i^a = 0$. Hence, like the LC connection $\Gamma^a_{b c}$, the
LC spin connection is invariant under
constant conformal transformations.
Secondly, the form in \eqref{var_LC_spin_connex} implies that $E_k^c \delta \Gamma_{c}^{k}$ is a
total divergence, given specifically by
\begin{equation}\label{tot_div}
E_k^c \delta \Gamma_{c}^{k}
=
\frac{1}{2}\,(\epsilon^{}_{ijk}
E_{a}^i
E_j^b
\delta E^a_{k})^{}_{,b} .
\end{equation}
One therefore has
$\Gamma_a^i(x) = \delta \F/\delta E_i^a(x)$
in terms of the ``generating function''
$\F := \int \Gamma_a^i E_i^a \d^3 x$, together with the
following integrability
identity \cite{Thiemann2001}:
\begin{equation}\label{fd_spinconn_int}
\fd{\Gamma_{a}^{i}(x)}{E^b_j(x')}
=
\fd{\Gamma_{b}^{j}(x')}{E^a_i(x)} .
\end{equation}

\textbf{Spin gauge formalism.}
By virtue of
\eqref{fd_spinconn_int},
a one-parameter family
of phase spaces of GR
can be constructed as:
\begin{align}\label{A^i_a}
A^i_a &:= \Gamma^i_a + \beta \, K^i_a,\;\; P_i^a := \frac{E_i^a}{\beta}
\end{align}
parametrized by
a nonzero complex constant
$\beta$. We refer to the pairs $(A^i_a, P_i^a)$
as the ``$A$-variables'' and denote the corresponding
PB by $\PBA{\centerdot\,}{\centerdot}$,
The transformation from the
$K$- to $A$-variables is canonical since the
following equations
%form a canonical set of variables for GR. They will be called the
%$A$-variables. We will consider $\beta$ as positive so that $P_i^a$
%has the same orientation as $E_i^a$.
%By virtue of \eqref{fd_spinconn_int}
%and inverting \eqref{A^i_a} and \eqref{P_i^a},
%we see that
\begin{align}
\PBA {K^i_a(x)}{E^b_j(x')}
&=
\delta^i_j \delta_a^b \,  \delta(x, x')
\label{PBAKE}
\\
\PBA {K^i_a(x)}{K^j_b(x')} &=
0
=
\PBA {E^a_i(x)}{E^b_j(x')}
\label{PBAKKEE}
\end{align}
hold for any $\beta$. In validating the first equation in \eqref{PBAKKEE},
the identity \eqref{fd_spinconn_int} has been evoked. It follows that
all PB relations in the $K$-variables are strongly preserved
in the $A$-variables. By design of
relations in \eqref{A^i_a}, $A^j_a$
is a spin connection with
$\tilde{K}^i_a := A^i_a - \Gamma^i_a = \beta K^i_a$
as the torsion contribution.
The associated spin
covariant derivative and curvature 2-form
are denoted by $D$ and $F^{k}_{a b}$ respectively, so that
any spin-valued scalar $S^i$ we have
\begin{align}
\label{spindelD}
&D_a S^i
=
\partial^{}_a S^i + \epsilon^{}_{i j k} A^j_a S^k
\\
&[D_a, D_b] S^i
=
\epsilon^{}_{i j k} F^j_{a b}\,S^k .
\end{align}
By using \eqref{torfree} and \eqref{PBAKE} we see that
$\C^K_i = \epsilon_{i j k} \, K_a^j \, E_k^a = P^a_{i,a} + \epsilon_{i j k} \, A_a^j \, P_k^a$
which allows one to express the spin
constraint in the form
\begin{equation}\label{Gk}
\C^A_i
:=
D_a P^a_i
\end{equation}
analogous to the ``Gauss constraint'' in the Maxwell and Yang-Mills gauge theories.
Furthermore, the momentum constraint becomes
\begin{align}
\H_a
&=
\C^A_a
+
\Gamma^k_{a} \C^A_k
-\frac12\,
\epsilon^{}_{ijk}\,E^i_a   E^b_j\, \C^{}_{k;b}
\\
\label{C^A_a}
\C^A_a
&:=
F^k_{ab} P_k^b
-
A^k_{a} \C^A_k .
\end{align}
The constraints $\C^A_k$ and $\C^A_a$ respectively
generates rotations and diffeomorphisms through their
PBs with the $A$-variables.
The Hamiltonian constraint then becomes:
\begin{align}\label{HHA}
\H_\perp
&=
\C_\perp^A
+
\frac{2\beta^2}{\mu}\,  P_k^c \C^{}_{k;c}
+
\frac1{8\mu}\,\C_{k}\C_{k}
\\
\C_\perp^A
&:=
\frac{1}{\mu}
\left[
\epsilon_{i j k}\,
\beta^2 F^k_{a b} -\frac{4\beta^2+1}{2}\,\tilde{K}^i_{[a} \tilde{K}^j_{b]}
\right]
P^a_i
P^b_j .
\label{HA}
\end{align}
%In this context, even through
%$\Kappa^i_{a}$ may be evaluated as $\beta K^i_a$, it is really
%$\beta$-independent.
Note that
terms like $P_k^c \C^{}_{k;c}$ in \eqref{HHA}
are regarded as proportional to the constraints
involved, as the LC covariant differentiation of the constraint there will
be swapped over to their coefficients once ``smeared'' over the
spatial hypersurface.
The constraints $\C^A_k, \C_a^A$ and  $\C_\perp^A$ hence form a set of independent
first class constraints.
Ashtekar's original gauge formalism of GR corresponds to
the choice $\beta=\pm i/2$ so that the non-polynomial term in \eqref{HA} vanishes.
In Barbero's modified approach,
$\beta$ is considered as a real and positive parameter in order
to resolve the reality problem on quantization.
%For real $\beta$, the orientation of $P^a_i$ is preserved for $\beta>0$.

\textbf{Conformal triad formalism.} We shall develop a new
set of real gauge variables for GR that contains no free
parameters. The ambiguity due to $\beta$ described above arises
from an arbitrary scaling factor in defining the $A$-variables. If
an alternative set of gauge variables for GR can be found that
possesses a conformal symmetry, then such an arbitrariness
may be absorbed. Therefore, the search for such variables
naturally involves an extension of the GR phase space by
incorporating this symmetry. By analogy with the passage
from the triad to spin gauge formalism of GR discussed above, we
shall first introduce a set of ``conformal triad variables'' as a
precursor of the ultimate spin gauge variables of GR with
conformal symmetry.

To this end, we
introduce the conformal triad $\bar{e}^i_a$ with inverse $\bar{e}_i^a$ so that
$\gamma_{a b} = \bar{e}^i_a \bar{e}^i_b$ and
$\gamma^{a b} = \bar{e}_i^a \bar{e}_i^b$.
%\begin{align}
%\label{epsilon_i^a}
%\bar{e}_i^a &= \phi^{2} \,e_i^a
%\\
%\label{epsilon^i_a}
%\bar{e}^i_a &= \phi^{-2} \,e^i_a
%\end{align}
Further, we introduce
the densitized triad $\Epsilon_i^a = \bar{\mu}\, \bar{e}_i^a$ with inverse
$\Epsilon_i^a = \bar{\mu}^{-1}\, \bar{e}^i_a$.
The LC spin connection of $\gamma_{a b}$ will be
denoted by $\bar{\nabla}$ and
the associated covariant differentiation
also denoted by a subscript ``$|$''.
Relations in complete analogue with
\eqref{spindel}--\eqref{fd_spinconn_int}
hold with the substitutions
\begin{equation}\label{subs}
g_{a b} \rightarrow \gamma_{a b},\;
\Gamma^i_a \rightarrow \bar{\Gamma}^i_a,\;
\nabla  \rightarrow \bar{\nabla},\;
E_i^a \rightarrow \Epsilon_i^a .
\end{equation}
A trace-split
of the extrinsic curvature $K^i_a$ is then performed
in a conformally invariant manner. These considerations lead to
the spin version of \eqref{ggamma} as
\begin{align}
\label{Epsilon_i^a}
E_i^a &= \phi^{4}\, \Epsilon_i^a,
\;\;
%\label{Kappa^i_a}
K^i_a
=
\phi^{-4}\, \Kappa^i_a + \frac12\,\phi^2 \bar{\mu}\, \Epsilon^i_a \tau
\end{align}
where we have introduced $\Kappa^i_a$ to function as the ``conformal
extrinsic curvature''. We have thus arrived at a set of conformal triad
description of GR using
$(\Kappa^C, \Epsilon_C):=(\Kappa^i_a, \Epsilon^a_i; \tau, \mu)$, called
the ``$\Kappa$-variables''. Using the corresponding PB denoted by
$\PBKappa{\centerdot\,}{\centerdot}$, we can show that these variables are
indeed canonical, since \eqref{Epsilon_i^a} implies
\begin{align}
\label{PBKappaKE}
\PBKappa {K^i_a(x)}{E^b_j(x')}
&=
\delta^i_j \delta_a^b \,  \delta(x, x')
\\
\PBKappa {K^i_a(x)}{K^j_b(x')}
&=
0
= \PBKappa {E^a_i(x)}{E^b_j(x')} .
\label{PBKappaKKEE}
\end{align}
%Therefore all PB relations in the $K$-variables
%are strongly preserved in the equivalent
%PB relations in the $\Kappa$-variables.
%The $\Kappa$-variables are related to the $K$-variables by
From \eqref{Epsilon_i^a} one might see the factor
$\phi^4$ as a generalization of $\beta$.
However, instead of being a free parameter,
it is important to note that this factor depends on the canonical variables
which in turn have a conformal symmetry.
%In terms of the $\Kappa$-variables, we
%see that
%\begin{align}
%\C^\Gamma
%&=
%\frac{1}{2}\,
%\Epsilon^c_k \,\Kappa^k_{c}
%=:
%\C^{\Kappa}
%\\
%\label{gauss2Kappa}
%\C_i^K
%&=
%\epsilon^{}_{i j k}\, \Kappa^{}_{a j} \Epsilon^a_{k}
%=: \C^{\Kappa}_i
%\end{align}

%\begin{align}
%K_{ab}
%=
%\frac{1}{2} \, \phi^{-2}\, \bar{\mu}\,\Kappa^i_{(a} \Epsilon^i_{b)}
%+
%\frac{1}{4}\,\phi^{4}  \bar{\mu}^2 \tau \,\Epsilon^i_{a} \Epsilon^i_{b}
%\label{K_ab}
%\end{align}

%\begin{align}
%\pi^{ab}
%&=
%\frac{1}{2}\,
%\bar{\mu}^{-2} \,
%\Epsilon^{a}_i \Epsilon^{b}_j
%\,
%\Kappa^{}_{c(i} \Epsilon^c_{j)}
%\label{pi2K}
%\end{align}

\textbf{Conformal spin gauge formalism.} Based on the preceding discussions,
we shall now formulate our final phase space of GR by incorporating
conformal symmetry while retaining a spin gauge structure.
The basis of the canonical transformation from the $K$- to $A$-variables
can be traced to the term $E_k^c \dot{\Gamma}_{c}^{k}$
being a total divergence due to \eqref{tot_div}. As such,
up to an arbitrary constant coefficient, say $1/\beta$,
it
can be added to the
time-derivative terms in the canonical action for GR
in the $K$-variables
as follows:
\begin{align}\label{}
E_i^a \dot{K}^i_a
+
\frac{1}{\beta}
E_k^c \dot{\Gamma}_{c}^{k} = P_i^a \dot{A}^i_a .
\end{align}
The
result is the time-derivative terms
in the $A$-variables.
In a similar fashion, the analogy of \eqref{tot_div}
with substitution \eqref{subs} enables us to
add the total divergence
$({1}/{\alpha})\Epsilon_k^c \dot{\bar{\Gamma}}_{c}^{k}$
to the time-derivative terms in the $\Kappa$-variables:
\begin{align}\label{EalKa}
\mu\,\dot{\tau}
+
\Epsilon_i^a\dot{\Kappa}^i_a
+
\frac{1}{\alpha}
\Epsilon_k^c \dot{\bar{\Gamma}}_{c}^{k}
=
\mu\,\dot{\tau}
+
\frac{\Epsilon_i^a}{\alpha}   (\bar{\Gamma}^i_a + \alpha\Kappa^i_a)\,\dot{}
\end{align}
for any constant ${1/\alpha}$. However,
owing to the ``built-in'' conformal symmetry of the
$\Kappa$-variables, this constant can always be absorbed
into the variables themselves using
\begin{equation}\label{}
\Epsilon_i^a\rightarrow \alpha \Epsilon_i^a,\;
\Kappa^i_a\rightarrow \frac{\Kappa^i_a}{\alpha},\;
\bar{\Gamma}^i_a\rightarrow \bar{\Gamma}^i_a
\end{equation}
so long as
$\alpha$ is real and positive. In this case,
\eqref{EalKa} yields the
terms
$\mu\,\dot{\tau}+\Pi_i^a\dot{\Alpha}^i_a$
in a new set of variables:
\begin{align}\label{Alpha^i_a}
\Alpha^i_a
&:=
\bar{\Gamma}^i_a + \Kappa^i_a,
\;\;
\Pi_i^a
:=
\Epsilon_i^a .
\end{align}
We call $(\Alpha^C, \Pi_C):=(\Alpha^i_a, \Pi_i^a; \tau, \mu)$
the ``$\Alpha$-variables'' and denote the
associated PB by $\PBAlpha{\centerdot\,}{\centerdot}$.
The canonical nature of these variables can be
verified by
the PB relations:
\begin{align}
\label{PBAKlphaE}
\PBAlpha {\Kappa^C(x)}{\Epsilon_D(x')}
&=
\delta^C_D \,  \delta(x, x')
\\
\PBAlpha{\Kappa^C(x)}{\Kappa^D(x')}
&=
0
=
\PBAlpha {\Epsilon_C(x)}{\Epsilon_D(x')}.
\label{PBAlphaKKEE}
\end{align}
In deriving the first equation in \eqref{PBAlphaKKEE}, we have
used the analogue of \eqref{fd_spinconn_int} with \eqref{subs}.
%Therefore all PB relations in the $K$-variables
%are strongly preserved in the equivalent
%PB relations in the $\Alpha$-variables.
The spin
covariant derivative
associated with $\Alpha^i_a$
and its curvature 2-form are
denoted by $\bar{D}$ and
$\bar{F}$ respectively.
Henceforth, $\Kappa^i_{a}$ is understood in terms of the $\Alpha$-variables
as $\Alpha^i_a - \bar{\Gamma}^i_a$.
It follows that
\begin{align}
\C^\Gamma
&=
\frac{1}{2}\,
\Kappa^i_{a} \Pi^a_i
=:
\C^{\Alpha}
\\
\label{gauss2Kappa}
\C_i^K
&=
\epsilon^{}_{i j k}\, \Kappa^{}_{a j} \Epsilon^a_{k}
=
\bar{D}_a \Pi^a_i
=: \C^{\Alpha}_i
\\
\H_a
&=
\C^\Alpha_a
+
\bar{\Gamma}^k_{a} \C^\Alpha_k
-\frac12
\epsilon^{}_{i j k}
\Pi^i_a \Pi^{b}_j
\C^\Alpha_{k|b}
-
2 \C^\Alpha_{|a}
+
4(\ln\phi)_{,a} \C^\Alpha
\nonumber
%\label{HaKappa}
\\
\C^\Alpha_a
&:=
\tau_{,a} {\mu}
+
\bar{F}^k_{ab} \Pi_k^b
-
\Alpha^k_{a} \C^\Alpha_k .
\label{CaKappa}
\end{align}
The constraints $\C^{\Alpha}$, $\C^{\Alpha}_i$ and
$\C^\Alpha_a$ may be called the
conformal, spin and diffeomorphism constraints
in the $\Alpha$-variables respectively,
as they generate the corresponding transformations
using $\PBAlpha{\centerdot\,}{\centerdot}$.
Finally, the Hamiltonian constraint in the form of \eqref{HR} with \eqref{H^Gamma} becomes
\begin{align}
\H_\perp
&=
\C_\perp^\Alpha
+
\frac{1}{8\mu}
\C^\Alpha_k\C^\Alpha_k
+
\frac{2\phi^8}{\mu}
\Pi_k^c \C^\Alpha_{k|c}
+
\frac\tau2\,\C^\Alpha
+
\frac{1}{2\mu}\,(\C^\Alpha)^2
\label{HRKappa1}
\\
\C_\perp^\Alpha
&:=
-
\frac{3}{8}\, \tau^2 \mu
+
8\mu\,
\phi^{-5} \bar{\Delta} \phi
\nonumber
\\
&
\;\;\;\;\:+
\frac{1}{\mu}
\left[
\phi^8
\epsilon_{kij}
\bar{F}^k_{ab}
-
\frac{4\phi^8+1}{2}
\Kappa^i_{[a}  \Kappa^j_{b]}
\right]
\Pi_i^a \Pi_j^b
\label{HAlpha}
\end{align}
where $\bar{\Delta} := \gamma^{ab}\bar{\nabla}_a\bar{\nabla}_b$ is
the Laplacian associated with the conformal metric $\gamma_{ab}$.
In \eqref{HAlpha}, the third term  is analogous to \eqref{HA} with
$\beta\rightarrow \phi^4$. There, the ``additional'' first term is
due to the York time $\tau$ being separated from the conformal
part of kinematics whereas the second term counts for the
conformal factor $\phi$ being a local function of the
$\Alpha$-variables.

By virtue of \eqref{HRKappa1} and the preservation of the PB of
$\H_\perp(x)$ and $\H_\perp(x')$ throughout all canonical
transformations considered in this work, we conclude that
$\C_\perp^\Alpha$ and the canonical generators $\C^{\Alpha}$,
$\C^{\Alpha}_i$ and $\C^\Alpha_a$ form a set of first class
constraints for the above conformal spin gauge formulation of GR
using the $\Alpha$-variables. There are no free parameters
in the description. The main tradeoff seems to be the two
extra leading terms in the effective Hamiltonian constraint
in \eqref{HAlpha}. While the first term looks quite
simple, the second term may post new challenges in the
regularization procedure on quantization. Nevertheless, the
absence of any parameter in the presented formulation of GR opens
up an interesting possibility for ``conformal loop quantum
gravity'' that is at least free from the Barbero-Immirzi
ambiguity. Furthermore, a unitary evolution of loop quantum
gravity may also be addressed as per the discussion towards the
end of \cite{Wang2005b}. Discussions on the related quantum
issues as well as the full detail of the present work are deferred
to a future publication.

%2nd terms has a structure analogous to $\H$ in the $\Alpha$ vars.
%However, still somehow more complicated. This seems the trade off
%of this parameter-free formulation. Further research is required
%regarding the quantum implementation of this formalism before a
%judgement may be made to see whether this apparent increase of
%complexity is merely of technical nature.

%\textbf{Remarks.}
%We have analyzed and compared different canonical formulism of GR
%with different extended symmetry properties ... Worked through a
%hierarchy of (PB-preserving) canonical transformations. Through a
%sequence of canonical transformations, we have finally arrived at
%... Strong indication that ... may lead to a unitary evolution of
%quantum gravity as per discussions towards the end of
%\cite{Wang2005b}.

%{\bf Acknowledgments.}
I thank A. E. Fischer and  C. J. Isham for
stimulating interactions and the CCLRC Centre
for Fundamental Physics for partial support.

\end{document}